\documentstyle[aps,12pt]{revtex}

\begin{document}
\title{Quantum Critical Point in Strongly Correlated $^{87}${\rm Rb} Atoms in
Optical Lattice}
\author{Su-Peng Kou$^{1,2}$ and Rong-Hua Li$^2$}
\address{$^1${Department of Physics, Massachusetts Institute of Technology, Cambridge}%
\\
MA 02139, USA\\
$^2$Department of Physics, Beijing Normal University, Beijing, 100875, China%
\vskip 0.5cm}
\maketitle

\begin{abstract}
\begin{center}
{\large Abstract}
\end{center}

In this paper, the Bosonic projected variational method is proposed to study
the strongly correlated $^{87}{\rm Rb}$ atoms in optical lattice. A global
phase diagram is obtained by this method. There exist two characteristic
lattice depths $V$ for $^{87}{\rm Rb}$ atoms in optical lattice : one is $%
V=9.5E_r$ to label the maximum height of the 'zero-momentum' peak of
condensation, the other is the quantum critical point for the
superfluid-insulator (SI) transition at $V=12.3E_r$. As a result of strongly
correlated effect for lattice Bosons, the suppressed superfluid state is
predicted near the SI transition with the suppressed superfluid density and
the very slowly velocity of the sound-like excitons.

\vskip 1cm

PACS (numbers): 03.75.Kk, 03.75.Lm, 03.75.Mn, 32.80.Pj

\vskip 1cm

Keywords: Bose-Hubbard model, superfluid-insulator transition; bosonic
partial projection operator

\newpage
\end{abstract}

During the last years a spectacular development in the storage and
manipulation of cold atoms in optical lattices has taken place. An
superfluid-insulator (SI) transition was observed for $^{87}{\rm
Rb}$ atoms trapped in a three-dimensional optical lattice
potential by changing the potential depth \cite{Greiner}. The SI
transition is the appearance of a excitation gap. In the
superfluid regime, there is no excitation gap and instead one
observes a series of 'Bragg-peaks' around the characteristic
'zero-momentum' peak of a condensate in the absence of an optical
lattice. The existence of this gap has been verified
experimentally by applying a phase gradient in the Mott-insulator.
This experimental progress has revived the interest in the
Bose-Hubbard (BH) model as a generic Hamiltonian for strongly
correlated bosons \cite{Fisher}, by which the quantum phase
transition can be described {\cite
{Sachdev,Jaksch,bru,she,rok,kra,sca,niy,bat,fre,zig,gor,stoof,ami,els,liu}}.

In this paper we intend to examine the physics in $^{87}{\rm Rb}$ atoms by
studying a Bose-Hubbard model using a variational method for bosonic systems~%
\cite{gutzwiller}. In the new approach, the on-site repulsion is treated
exactly, while the kinetic energy is studied variationally, so that it is
suitable to examine some issues in strongly correlated Bosonic systems. The
variational method applied to the Bose-Hubbard model in three-dimension
demonstrates a quantum phase transition from a superfluid with suppressed
superfluid density for smaller intra-site Coulomb repulsion $U$ to a Mott
insulator for larger $U$ at unit filling $\frac NL=1$ . There exists a
quantum critical point (QCP) for the homogenous phase. We calculate the
superfluid density near the QCP and show the existence of the suppressed
superfluid state which is a kind of ''gossamer'' phenomenon. The idea of
''gossamer'' (superconducting) state is that the ``insulator'' might
actually be a thin, ghostly superconductor which is proposed by Laughlin\cite
{laugh,zhang}. In a suppressed superfluid state, the superfluid density is
very thin and the velocity of the sound-like excitons is very slow, in
contrast to the conventional superfluid state.

The system is based on confining cold $^{87}{\rm Rb}$ atoms in the periodic
potential of an optical lattice\cite{Greiner}. In the simplest case, three
orthogonal, independent standing laser fields with wave vector $k$ produce a
separable three dimensional lattice potential $V(x,y,z)=V\left( \sin ^2{kx}%
+\sin ^2{ky}+\sin ^2{kz}\right) $ with a tunable amplitude $V\gg E_r=\hbar
^2k^2/2m$. Starting from the standard pseudopotential description the
interatomic potential is replaced by an effective contact interaction of the
form $U(\vec{x})=\frac{4\pi \hbar ^2a_s}m\cdot \delta (\vec{x})$ containing
the exact s-wave scattering length $a_s$. With $\hat{a}_i^{\dagger }$ as the
creation operator of a boson at site $i$ and $\hat{n}_i$ as the density
operator, the Hamiltonian reads

\begin{equation}
\hat{H}=-J\sum_{<ij>}\,\hat{a}_i^{\dagger }\hat{a}_j\,+\,\frac U2\sum_i\,%
\hat{n}_i(\hat{n}_i-1)
\end{equation}
The bandwidth parameter $J$ is essentially the gain in kinetic energy due to
nearest neighbor tunneling $J=\frac 4{\sqrt{\pi }}E_r\left( \frac V{E_r}%
\right) ^{3/4}\exp {-2\left( \frac V{E_r}\right) ^{1/2}}.$ The relevant
interaction parameter $U$ is thus given by an integral over the on-site wave
function via $U=\sqrt{\frac 8\pi }ka_sE_r\left( \frac V{E_r}\right) ^{3/4}.$

We consider $\mid \Psi \rangle =\hat{P}_p\mid \Phi _0\rangle $ as
a variational wavefunction to examine the ground state of the
above effective BH model for $^{87}{\rm Rb}$ atoms in optical
lattices. $\mid \Phi _0\rangle $ is the wavefunction of the
Bose-condensed ground state for non-interacting Bosons
\begin{equation}
|\Phi _0>=\exp (\sqrt{N}e^{i\varphi _0}\hat{a}_{{\bf k=0}}^{\dagger })|0>
\end{equation}
where $N$ is the number of Bosons. The order parameter $\langle \Phi _0|\hat{%
a}_{\vec{k}=0}|\Phi _0\rangle =e^{i\varphi _0}\sqrt{N}$ shows an off
diagonal long range order. $\hat{P}_p$ is the Bosonic partial projection
(BPP) operator defined as
\begin{equation}
\hat{P}_p=\prod_{m=2}^N\prod_i[1-\prod_{j=0,j\neq m}^N\frac{(\hat{N}_i-j)}{%
(m-j)}\cdot (1-\varepsilon _m)].
\end{equation}
If the on-site repulsive interaction between bosons is large enough, there
exist little possibly of high occupation states, we can introduce the
Bosonic partial projection (BPP) operator to describe the strongly
correlated effects. If $\varepsilon _m=0,$ the Bosonic partial projection
operator $\hat{P}_p$ turns into the Bosonic completely projection operator $%
\hat{P}_c,$ the ground state $\mid \Psi \rangle $ is reduced into a Mott
insulator state with gapped excitons. If $\varepsilon _m=1,$ there is no
interaction between Bosons, the Bosonic partial projection operator $\hat{P}%
_p$ turns into constant number $\hat{P}_p=1,$ and the ground state turns
into a Bose-condensation state with massless excitons $E\sim \alpha q^2$.

The variational energy for the ground state becomes
\begin{eqnarray}
E_g &=&\langle \Psi \mid \hat{H}\mid \Psi \rangle /\langle \Psi \mid \Psi
\rangle \\
&=&\langle \hat{H}_t\rangle +UD_2+3UD_3+...+\frac{N(N-1)}2UD_N.  \nonumber
\end{eqnarray}
The occupation numbers are $D_2$, $D_3,$ ...$D_N$. The variational kinetic
energy $\langle H_t\rangle $ is
\begin{equation}
\langle \hat{H}_t\rangle =\langle \Psi \mid -J\sum\limits_{\langle ij\rangle
}(\hat{a}_i^{\dag }\hat{a}_j+h.c.)\mid \Psi \rangle /\langle \Psi \mid \Psi
\rangle =g_JE_0
\end{equation}
where $E_0$ the kinetic energy for non-interaction Bose systems
\begin{equation}
E_0=\langle \Phi _0\mid -J\sum\limits_{\langle ij\rangle }(\hat{a}_i^{\dag }%
\hat{a}_j+h.c.)\mid \Phi _0\rangle /\langle \Phi _0\mid \Phi _0\rangle .
\end{equation}
$g_J$ is renormalization factor for kinetic energy to be determined.

In the thermodynamics limit $N\rightarrow \infty $, we choose the maximum
term of the variational wavefunction $\mid \Psi \rangle $ and obtain the
condition
\begin{eqnarray}
\varepsilon _m^2 &=&\frac{d_m(x+d_2+2d_3+...+(N-1)d_N)^{m-1}}{%
(1-x-2d_2-3d_3...-Nd_N)^m}, \\
\text{ }m &=&2,3,...N.  \nonumber
\end{eqnarray}
where $x=(N-D_2-D_3-...D_N)/L$ describes the empty concentration. $L$ is the
number of the site. From the condition we have the renormalization factor
for the kinetic energy
\begin{eqnarray}
g_J &=&\sum_{m=1}^NC_m\frac{A^m}{B^{m-2}}, \\
C_m &=&\varepsilon _{m-1}\varepsilon _m+\varepsilon _2\varepsilon
_{m-2}\varepsilon _{m-1}+\varepsilon _2\varepsilon _3\varepsilon
_{m-3}\varepsilon _{m-2}+...  \nonumber \\
&&+\varepsilon _i\varepsilon _{m-i}\varepsilon _{i-1}\varepsilon
_{m-i+1}+...+\varepsilon _{m-1}\varepsilon _m.  \nonumber
\end{eqnarray}
with $A^m=1-x-2d_2-3d_3...-Nd_N$ and $B=x+d_2+2d_3+...+(N-1)d_N.$

In the strong interaction limit, one has $\varepsilon _m^2\simeq \frac{%
d_m(x+d_2)^{m-1}}{(1-x-2d_2)^m}.$ The renormalization factor for the kinetic
energy is obtained as
\begin{eqnarray}
g_J &\simeq &(d_2+x)(1-x-2d_2) \\
&&+...2\sqrt{d_{m-2}d_{m-1}}[(d_2+x)(1-x-2d_2)]^{1/2}+....  \nonumber
\end{eqnarray}
where $d_m=\frac{D_m}L$ are determined by the equations
\[
\frac{\partial E_g(d_m)}{\partial d_m}=0.
\]
In this paper we solve the equations by the approximation up to $d_2$
without considering $d_m,$ $m>2$\cite{1}. The renormalization factor is
obtained as $g_J\approx (1-x-2d_2)(2d_2+x+2\sqrt{d_2(d_2+x)}).$ $d_2$ is
determined by the minimum energy condition $\frac{\partial E_g}{\partial d_2}%
=0\ $as $d_2\simeq \frac 14[1-(\frac U{24J})].$ In the following paper we
consider a BH model with unit filling $N/L=1$ or $x=0$.

Note that in the BPP approach and at unit filling case $N/L=1$ or $x=0$, $%
d_2 $ is a measure of the mobile carrier density.{\em \ }At $d_2=0$, we have
$\langle H_t\rangle =0$. This state describes a Mott insulator. On the other
hand, the case with $d_2>0$ describes superfluid state. We expect a
transition from the Mott insulator at larger $U$ to the superfluid at
smaller $U$ as $U$ decreases passing through a critical point $U_c$. The
transition point $U_c$ is given by
\begin{equation}
U_c=(\frac{-\partial g_J}{\partial d_2})_{\mid _{d_2=0}}\frac{E_0}N=24J
\end{equation}
or $({\frac V{E_r})}_c=\frac 14\ln ^2(\frac{\sqrt{2}ka_s}{48})\simeq 12.3.$
For $U>U_c=24J$, there is no solution for physical values of $d_2$,
indicating that $d_2=0$.

If $U\leq U_c$ $($ $({\frac V{E_r})}_c\leq 12.3)$, $d_2>0$ $(\varepsilon
_2>0),$ the ground state is a superfluid state.

Within the BPP approximation, the suppressed superfluid density $\langle
\Phi |\hat{n}_{\vec{k}=0}|\Phi \rangle $ is
\begin{equation}
\frac{\langle \Phi |\hat{n}_{\vec{k}=0}|\Phi \rangle }{\langle \Phi _0|\hat{n%
}_{\vec{k}=0}|\Phi _0\rangle }=g_J=\frac 12[1-(\frac U{U_c})^2].
\end{equation}
In this state, the mobile carrier density $d_2$ is fixed and the excitons
are the phase fluctuations - phasons. To consider the dynamics of phasons,
the quantum states are defined as
\begin{equation}
|\tilde{\Psi}>=\hat{P}_p\exp (\sum_j^Le^{i\varphi _i}a_j^{+})|0>.
\end{equation}
The effective Hamiltonian becomes
\begin{eqnarray}
H_{eff} &=&\langle \tilde{\Psi}\mid -J\sum\limits_{\langle ij\rangle }(\hat{a%
}_i^{\dag }\hat{a}_j+h.c.)+\frac{{U}}2\sum_in_i(n_i-1)\mid \tilde{\Psi}%
\rangle /\langle \tilde{\Psi}\mid \tilde{\Psi}\rangle \\
&\sim &-\rho _p\sum\limits_{\langle ij\rangle }\cos (\varphi _i-\varphi _j)+{%
U\sum_i(}\frac d{d\varphi _i})^2,  \nonumber
\end{eqnarray}
where $\rho _p=\frac J4[1-(\frac U{U_c})^2]$ is the phase stiffness. As a
result the effective model of the SF state turns into a quantum four
dimensional XY model with long range order. The transition temperature $T_c$
is scaled as $T_c\sim \rho _p^{2/3}$\cite{Fisher}.

In the SF regime, there is no excitation gap and instead the homogeneous
system exhibits a sound like mode - phason with frequency $\omega (q)=cq.$
The associated sound velocity is $c=\sqrt{\frac 14JU[1-(\frac U{U_c})^2]}$.
It is the correlated effect drives the BEC-like $q^2$ spectrum to
superfluid-like $q$ type.

The SF phase of $^{87}${\rm Rb} atoms in an optical lattice can thus quite
generally be characterized by the fact that at reciprocal lattice vectors $%
\vec{q}=0$, the momentum distribution $n_0$ has a peak with the height as%
\cite{Greiner,rot}

\begin{eqnarray}
n_0 &=&L\cdot \langle \Phi |\hat{n}_{\vec{k}=0}|\Phi \rangle |w(0)|^2 \\
&\sim &L\frac 12[1-\sqrt{2}ka_s\exp (-\sqrt{4V/E_r})](\frac V{E_r})^{3/2}.
\nonumber
\end{eqnarray}
The fact that the peaks in the momentum distribution at $\vec{q}=0$
initially grow with increasing depth of the lattice potential is a result of
the strong decrease in spatial extent of the Wannier function $w(\vec{x})$,
which entails a corresponding increase in its Fourier transform $w(\vec{q}%
)\propto (\frac V{E_r})^{3/4}$. Beyond a critical lattice depth around $%
V=9.5E_r$, this trend is reversed, however, and the superfluid density
eventually disappear completely at $U=U_c$ $(V=12.3E_r)$.

If $U>U_c$ or $({\frac V{E_r})}_c>12.3,$ $d_2=0$ and $\varepsilon _2=0$ ( $%
d_m=0$ and $\varepsilon _m=0,$ $m>2$)$,$ the ground state is a Mott
insulator, of which the wave function is
\begin{equation}
|\Phi >=\hat{P}_c|\Phi _0>=L^{-1/2}\prod\limits_l(|1_l>).
\end{equation}

The energy gap describes the energy difference between the ground state with
$D_2=0$ and the excited state with $D_2=1$ as $\Delta E\simeq -24J\cos
(aq/2)+U.$ The excited energy $\Delta E$ is from $(U-24J)$ to $(U+24J)$ with
the center at $U.$ $\Delta $ is the Mott gap for Boson exciton defined as $%
\Delta =(U-24J)$. Deep in the MI phase, this gap has size $U$, which is just
the increase in energy if an atom tunnels to an already occupied adjacent
site. The existence of this gap has been verified experimentally by applying
a phase gradient in the MI and measuring the resulting excitations produced
in the SF at smaller $V/E_r$\cite{Greiner}. Near the Mott transition, the
gap closes $\Delta =(U-24J)=0.$ The signal of the Mott gap $\Delta $
disappeared around $V=12.3E_r$ $(U=24J),$ which was taken also as another
definition of the critical point of the SI transition.

Above results predict two characteristic lattice depths: $V=9.5E_r$ and $%
V=12.3E_r$. Near the lattice depth around $V=9.5E_r$, the 'zero-momentum'
peak has maximum height at which the condensation is most robust. Beyond
this point, the phase coherence becomes weak. The other character lattice
depth is $V=12.3E_r$ which is just the QCP for the SI transition. At this
point the 'zero-momentum' peak disappears together with the opening of a gap
for particle-hole excitons. From the experiments by M. Greiner, at almost
the same point $V\simeq 13E_r$ the 'zero-momentum' peak of a condensate has
maximum height and the Mott gap $\Delta $ opens. Thus there exist only one
characteristic lattice depth$.$ The difference between our results and the
experiments is due to the inhomogeneous\cite{Greiner,Jaksch,Batrouni}. In
this paper our theory is based on the homogenous phase without considering
the inhomogeneous phase. The inhomogeneous phase in the BH model from BPP
method will be explored elsewhere.

In strongly{\em \ }correlated limit near the SI critical point $0<\frac{U-U_c%
}{U_c}\ll 1$, a new class of superfluid state appears - the suppressed
superfluid state. The wavefunction for so called the suppressed superfluid
state is read as
\begin{eqnarray}
|\Phi &>&=\hat{P}_p|\Phi
_0>=L^{-1/2}\prod\limits_l(|0_l>+|1_l>+\sum_{n_l=1}^N\frac{\varepsilon _{n_l}%
}{n_l!}|n_l>), \\
\text{ }\varepsilon _{n_l} &\simeq &\frac{d_m(x+d_2)^{m-1}}{(1-x-2d_2)^m}\ll
1.  \nonumber
\end{eqnarray}

It is obviously that the superfluid density for the suppressed superfluid is
suppressed seriously by correlations
\begin{equation}
n(\vec{q}=0)=\langle \Phi |\hat{n}_{\vec{q}=0}|\Phi \rangle =\frac 12(1-(%
\frac U{U_c})^2)\rightarrow 0,
\end{equation}
a quantitative measure of the suppressed superfluid. Another feature for the
suppressed superfluid is the existence of a very slowly velocity of the
phasons. The associated sound velocity turns into zero $c=\sqrt{\frac 14%
JU[1-(\frac U{U_c})^2]}\rightarrow 0$. The third character for the
suppressed superfluid is the pinned chemical potential at the center of the
gap.

Let us now consider the evolution of chemical potential. The chemical
potential $\mu $ is $\mu =\mu _0+\frac{\partial g_J}{\partial x}$. The term
originates from the $x$ dependences of $g_J$ in the variational procedure~%
\cite{zhang}, which will be important in calculation of the chemical
potential of the state
\begin{eqnarray}
\frac{\partial g_J}{\partial x} &=&1-2(2d_2+x)-2\sqrt{d_2(d_2+x)} \\
&&+(1-2d_2-x)\sqrt{(d_2+x)/d_2}.  \nonumber
\end{eqnarray}
In the limit $x\rightarrow 0,$ $U\rightarrow U_c,$ $\frac{\partial g_J}{%
\partial x}\sim 2-8d_2,$ one has
\begin{equation}
\mu \rightarrow \mu _0+\frac U2.
\end{equation}

Our results show that the suppressed superfluid state is a similar
phenomenon to the gossamer superconductivity. From the results of the
partial projection to the Gutzwiller variational method, the ground state
for strongly correlated electrons may be superconducting at half filling due
to some kinds of attraction mechanism\cite{zhang}. In this paper it is shown
that because of the partial projection, there exists the suppressed
superfluid state at unit filling $\frac NL=1$ in the region of $0<\frac{U-U_c%
}{U_c}\ll 1$. The Bose condensation can be suppressed seriously by strongly
correlated effect.

In summary, we have used the Bosonic projected variational method to study
the Bose-Hubbard model which is the effective model for strongly correlated $%
^{87}{\rm Rb}$ atoms in the optical lattice. And we obtain a global phase
diagram of the homogenous phase and give two results : two characteristic
lattice depths in phase diagram and the existence of the ''gossamer''-like
state near the SI transition.

\acknowledgements
The author thanks W.M. Liu for helpful conversations. Kou Su-Peng.
acknowledges support from NSFC Grant No. 10574014 and Beijing Normal
university.{}

\begin{center}
{\bf Figure Captions}
\end{center}
Fig.1

This figure shows the double occupation rate $d_2$ and the triple
occupation rate $d_3$ of $\gamma =\frac U{E_0}.$

Fig.2

Phase diagram for the Bose-Hubbard model under unit filling. In
this figure the scales at left axis and right axis are
$T_c(U\rightarrow 0)$ and $U_c$, respectively. $\frac{V_0}{E_r}$
is the parameter for Bosons in optical lattice.

\newpage

\begin{center}
{\Large References}
\end{center}

\begin{references}
\bibitem{Greiner}  M. Greiner, O. Mandel, T. Esslinger, T. W. Hansch, and I.
Bloch, Nature {\bf 415}, 39 (2002).

\bibitem{Fisher}  M.P.A. Fisher, P.B. Weichman, G. Grinstein, and D.S.
Fisher, Phys. Rev. B {\bf 40}, 546 (1989).

\bibitem{Sachdev}  {S. Sachdev, {\it Quantum Phase Transitions}, (Cambridge
Univ. Press, Cambridge, 2001).}

\bibitem{Jaksch}  {D. Jaksch, C. Bruder, J.~I. Cirac, C.~W. Gardiner, and P.
Zoller, Phys. Rev. Lett. {\bf 81}, 3108 (1998); D. Jaksch, V. Venturi, J. I.
Cirac, C. J. Williams, and P. Zoller Phys. Rev. Lett. 89, 040402 (2002). }

\bibitem{bru}  C. Bruder, R. Fazio, and G. Sch\"{o}n, Phys. Rev. B {\bf 47},
342 (1993). A.P. Kampf and G. T. Zimanyi, Phys. Rev. B {\bf 47} 279 (1993).
K.

\bibitem{she}  Sheshadri, H.R. Krishnamurty, R. Pandit, and T.V.
Ramakrishnan, Europhys. Lett. {\bf 22}, 257 (1992).

\bibitem{rok}  D.S. Rokhsar, B.G. Kotliar, Phys. Rev. B {\bf 44}, 10328
(1991).

\bibitem{kra}  W. Krauth, N. Trivedi, and D. Ceperley, Phys. Rev. Lett. {\bf %
67}, 2307 (1991). W. Krauth and N. Trivedi, Europys. Lett. {\bf 14}, 627
(1991); W. Krauth, M. Caffarel, and J.-P. Bouchard, Phys. Rev. B {\bf 45},
3137 (1992).

\bibitem{sca}  R.T. Scalettar, G.G. Batrouni and G.T. Zimanyi, Phys. Rev.
Lett. {\bf 66} 3144 (1991)

\bibitem{niy}  P. Niyaz, R.T. Scalettar, C.Y. Fong and G.G. Batrouni, Phys.
Rev. B {\bf 44}, 7143 (1991)

\bibitem{bat}  G.G. Batrouni and R.T. Scalettar, Phys. Rev. B {\bf 46}, 9051
(1992).

\bibitem{fre}  J.K. Freericks and H. Monien, Europhys. Lett. {\bf 26}, 545
(1993).

\bibitem{zig}  W. Zwerger, Journal of Optics B {\bf 5}, S9 (2003).

\bibitem{gor}  K. Goral, L. Santos and M. Lewenstein, Phys. Rev. Lett. {\bf %
88}, 170406 (2002).

\bibitem{stoof}  D. van Oosten, P. van der Straten, H.T.C. Stoof, Phys. Rev.
A {\bf 63}, 053601 (2001). D.B.M. Dickerscheid, D. van Oosten, P.J.H.
Denteneer and H.T.C. Stoof, Phys. Rev. A 68, 043623 (2003).

\bibitem{ami}  L. Amico and V. Penna, E-print: cond-mat/9801074.

\bibitem{els}  N. Elstner and H. Monien, E-print: cond-mat/9905367.

\bibitem{liu}  Jun-Jun Liang, J.-Q. Liang, and W.-M. Liu, Phys. Rev. A {\bf %
68}, 043605 (2003).

\bibitem{gutzwiller}  M.C. Gutzwiller, Phys. Rev. A {\bf 137}, 1726 (1965).

\bibitem{laugh}  R.B. Laughlin, E-print: cond-mat/0209269.

\bibitem{zhang}  F.C. Zhang, {Phys. Rev. Lett} {\bf 90}, 23 (2003).

\bibitem{1}  In strong interaction limit $U>>J$, $d_3$ is much smaller than $%
d_2$, $d_3\propto \frac 94d_2^2(1-2d_2)<<d_2.$

\bibitem{Kashurnikov}  V.A. Kashurnikov, N.V. Prokof'ev, B.V. Svistunov,
Phys. Rev{\it .} A {\bf 66} 031601(R) 2002.

\bibitem{rot}  R. Roth and K. Burnett 2002, {\it Preprint} cond-mat/0209066.

\bibitem{Batrouni}  G.G. Batrouni et al., Phys. Rev. Lett, {\bf 89}, 117203
(2002).
\end{references}
\end{document}